\documentclass[preprintnumbers,article,amsmath,amssymb,floatfix,10pt,prd,onecolumn,
superscriptaddress,nofootinbib]{revtex4-2}
\usepackage{bm}
\usepackage{amsfonts}
\usepackage{latexsym}
\usepackage[latin1]{inputenc}
\usepackage{graphicx}
\usepackage{amsmath}
\usepackage{palatino}
\usepackage{mathpazo}
\usepackage{textcomp}
\usepackage{float}
\usepackage{booktabs}
\usepackage{dcolumn}
\usepackage{ragged2e}
\usepackage{hyperref}
\hypersetup{colorlinks,citecolor=blue}
\hypersetup{colorlinks=true,linkcolor=red,filecolor=magenta,    urlcolor=blue}
\usepackage{xcolor}
\usepackage{orcidlink}
\usepackage{epsfig}
\usepackage{subfigure}
\usepackage{commath}

\def\jnl@style{\it}
\def\aaref@jnl#1{{\jnl@style#1}}

\def\aaref@jnl#1{{\jnl@style#1}}

\def\aj{\aaref@jnl{AJ}}                   
\def\apj{\aaref@jnl{ApJ}}                 
\def\apjl{\aaref@jnl{ApJ}}                
\def\apjs{\aaref@jnl{ApJS}}               
\def\apss{\aaref@jnl{Ap\&SS}}             
\def\aap{\aaref@jnl{A\&A}}                
\def\aapr{\aaref@jnl{A\&A~Rev.}}          
\def\aaps{\aaref@jnl{A\&AS}}              
\def\mnras{\aaref@jnl{Mon.~Not.~Roy.~Astron.~Soc.}}             
\def\prd{\aaref@jnl{Phys.~Rev.~D}}        
\def\prc{\aaref@jnl{Phys.~Rev.~C}}  
\def\prl{\aaref@jnl{Phys.~Rev.~Lett.}}    
\def\qjras{\aaref@jnl{QJRAS}}             
\def\skytel{\aaref@jnl{S\&T}}             
\def\ssr{\aaref@jnl{Space~Sci.~Rev.}}     
\def\zap{\aaref@jnl{ZAp}}                 
\def\nat{\aaref@jnl{Nature}}              
\def\aplett{\aaref@jnl{Astrophys.~Lett.}} 
\def\apspr{\aaref@jnl{Astrophys.~Space~Phys.~Res.}} 
\def\physrep{\aaref@jnl{Phys.~Rep.}}      
\def\physscr{\aaref@jnl{Phys.~Scr}}       
\def\commat{\aaref@jnl{Comm.~Math.~Phys.}}              
\def\science{\aaref@jnl{Science}}               
\def\cqg{\aaref@jnl{Classical Quant.~Grav.}}            
\def\jpcs{\aaref@jnl{JPCS}}                                     
\def\ijmpd{\aaref@jnl{Int.~J.~Mod.~Phys.~D}}                    
\def\grg{\aaref@jnl{Gen.~Relat.~Gravit.}}               
\def\rpp{\aaref@jnl{Rep.~Prog.~Phys.}}          
\def\npa{\aaref@jnl{Nucl.~Phys.~A}}        
\def\lrr{\aaref@jnl{Living Rev.~Rel.}}                   
\def\jcap{\aaref@jnl{J.~Cosmology Astropart.~Phys.}}    
\def\rmp{\aaref@jnl{Rev.~Mod.~Phys.}}   
\def\epjc{\aaref@jnl{Eur.~Phys.~J.~C}} 
\def\plb{\aaref@jnl{~Phy.~Lett.~B}} 
\def\mpla{\aaref@jnl{Mod.~Phy.~Lett.~A}} 
\def\arxiv{\aaref@jnl{arxiv.org}}

\allowdisplaybreaks[1]
\renewcommand{\arraystretch}{1.1}
\begin{document}

\color{black}      

\title{Wormhole solutions in $f(Q,T)$ gravity with a radial dependent B parameter}

\author{Moreshwar Tayde\orcidlink{0000-0002-3110-3411}}
\email{moreshwartayde@gmail.com}
\affiliation{Department of Mathematics, Birla Institute of Technology and
Science-Pilani,\\ Hyderabad Campus, Hyderabad-500078, India.}

\author{Joao R. L. Santos\orcidlink{0000-0002-9688-938X}}
\email{joaorafael@df.ufcg.edu.br}
\affiliation{UFCG - Universidade Federal de Campina Grande - Unidade Acad\^{e}mica de F\'isica,  58429-900 Campina Grande, PB, Brazil.}

\author{Julia N. Araujo \orcidlink{0000-0002-4684-0502}}
\email{julia.neves@estudante.ufcg.edu.br}
\affiliation{UFCG - Universidade Federal de Campina Grande - Unidade Acad\^{e}mica de F\'isica,  58429-900 Campina Grande, PB, Brazil.}

\author{P.K. Sahoo\orcidlink{0000-0003-2130-8832}}
\email{pksahoo@hyderabad.bits-pilani.ac.in}
\affiliation{Department of Mathematics, Birla Institute of Technology and
Science-Pilani,\\ Hyderabad Campus, Hyderabad-500078, India.}

\date{\today}

\begin{abstract}

 A possible astrophysical object to be found in General Relativity is the wormhole. This special solution describes a topological bridge connecting points in two distinguished universes or two different points in the same universe. Despite it was never observed so far, is desired to find traversable wormholes, i.e. wormholes which have a throat at which there is no horizon. However, the traversable wormhole constraints yield solutions that violate all the energy conditions in General Relativity. In the last few years, several models to describe gravity beyond $\Lambda$CDM have been proposed. Then, it is relevant to look for wormhole solutions for these new theories. In this study, we are going to unveil new wormhole solutions for the so-called $f(Q,T)$ gravity. This theory of gravity is based on the non-metricity scalar $Q$, which is responsible for the gravitational interaction together with the energy-momentum trace $T$. We use the embedding procedure to find both the energy and the equilibrium conditions for the existence of wormholes. Then, the nontrivial contributions coming from $f(Q,T)$ gravity are embedded into the effective equations for density and pressures. We also considered the presence of strange matter in the wormhole throat. Such a matter obeys the notorious MIT bag Model. We are going to present new scenarios confirming the viability of traversable wormholes in $f(Q,T)$ gravity with the strange matter, satisfying SEC and WEC energy conditions.
\end{abstract}


\maketitle

\textbf{Keywords:} $f(Q,T)$ gravity, wormhole, energy conditions, strange matter, MIT bag model

\section{Introduction}

The present high-level experiments on astrophysics and cosmology are pushing us toward a new era of discoveries about our Universe. Collaborations such as LIGO (Laser Interferometer Gravitational-Wave Observatory) \cite{Gourgoulhon/2019}, Virgo \cite{Abuter/2020}, Event Horizon Telescope (EHT) \cite{Chael/2016, Akiyama/2019}, International Gamma-Ray Astrophysics Laboratory (INTEGRAL) \cite{Winkler/2003}, Advanced Telescope for High-Energy Astrophysics (ATHENA) \cite{Barcons/2017}, Imaging x-ray Polarimetry mission (IXPE) \cite{Soffitta/2013}, Swift \cite{Burrows/2005}, CHIME \cite{Chime/2018} have been testing gravity and astrophysical objects as we never saw before. Moreover, new surveys such as LISA \cite{Lisa/2013}, BINGO \cite{Abdalla/2021} and SKA \cite{Hall/2004} have the potential to impose really strong boundaries on gravity theories, restricting the huge amount of theoretical proposals which have been appearing in the literature so far. 

An intriguing and exotic type of solution of Einstein's equations for General Relativity is the wormhole \cite{Visser}. The wormhole can be though as a topological bridge connecting points in two distinguished universes or two different points in the same universe. Despite been a theoretical solution, there are several studies on the viability of finding wormholes. An interesting route to find observable wormholes was proposed by Bueno et al. \cite{Bueno/2018}, there the authors use gravitational waves measurements to study echoes of the gravitational wave signal at the horizon scale of black holes. These gravitational waves would be connected to the postmerger ringdown phase in binary coalescences. Moreover,  Paul et al. \cite{Paul/2020}, worked on distinctive features of the ahasccretion disk images to distinguish between a wormhole geometry and a black hole, unveiling another interesting path for observation. Also, gravitational lensing by a wormhole and light deflection has been widely investigated in \cite{Ovgun/2019,Ovgun/2018,Ovgun/2020}.

Another desired features relative to wormholes is the search for traversable solutions, which means, wormholes which are big enough, in such a way that a person could traverse them and survive the tidal forces. This class of wormholes was introduced in the seminal paper of Morris and Thorne \cite{Morris/1988} and is still a hot subject, as we can see in the beautiful work of Maldacena and Milekhin \cite{Maldacena/2021}. There the authors presented a new dark sector based on the Randall-Sundrum model \cite{RS/1999} for a $U(1)$ gauge field which interacts with Standard Model particles only through gravity, and where traversable wormholes could exist. 

The main problem with traversable wormholes is their violation of all energy conditions in General Relativity. Consequently,  there is a need for searching for traversable wormhole solutions in different theories of gravity that could obey energy conditions. Another interesting issue is the matter inside the wormhole throats. If Null and Dominant energy conditions are violated, we have the presence of exotic matter in this region. In this work, we are going to connect this exotic matter with the strange matter distribution of the famous MIT Bag Model.  This model, originally presented by \cite{Chodos/74}, was motivated to understand the biding of strange quark matter. Later such a model was applied in the context of astrophysical objects as pointed out by Witten \cite{Witten}. There, it is highlighted that if one considers a star made of quark matter and ignores the contributions due to their masses, the relation between pressure and density is $p=\frac{1}{3}(\rho - 4B)$, where the vacuum pressure $B$ on the Bag wall is responsible to stabilize the confinement of the quarks. 

 So far, the General Relativity (GR) is the widely accepted theory of gravity, and it has been supported by several experiments and observations. However, there are some phenomena that General Relativity cannot entirely account for, such as the observed acceleration of the expansion of the Universe, the behavior of gravity on galactic scales, and the search for a quantum description of gravity. In response to these problems, alternative theories of gravity have been proposed. These theories modify or extend General Relativity in various ways. One broadly example of extension of GR is the so-called $f(R)$ gravity. This  theory modifies the Einstein-Hilbert action replacing the Ricci scalar by a function of the scalar curvature \cite{Hans A. Buchdahl,A. A. Starobinsky,A. D. Felice}. This modification leads to changes in the gravitational field equations and can affect the behavior of gravity at different scales. This theory has been used to explain the observed acceleration of the expansion of the Universe, corroborates with bounds over the primordial inflation and bypass the need for dark matter to explain the galaxies rotation curves data, stellar dynamics and galaxy morphology \cite{Capozziello/2012}. 

In 2011, Harko et al. \cite{Harko/2011} introduced an extension of $f(R)$ called $f(R,T)$ gravity. This theory of gravity consists in add an extra contribution to the Einstein's equations due to the trace of the energy-momentum tensor. Such a contribution would account for the classical manifestation of non-trivial quantum effects related with conformal anomaly aspects of the theory. This theory of gravity has been tested in several phenomenological and theoretical approaches as one can see in \cite{Sardar:2023, Pappas:2022, Bose:2022, joaof}.  

Besides the previous extensions of GR we also highlight the $f(Q)$ gravity, introduced by Jimenez et al. \cite{Jimenez}.  In such a theory, the non-metricity scalar $Q$ is responsible for the gravitational interaction. The $f(Q)$ gravity has been tested by several observational data in the last few years, as we can see in the work of Lazkoz et al. \cite{Lazkoz/2019}. There the authors constrained $f(Q)$ gravity using data from the expansion rate, Type Ia Supernovae, Quasars, Gamma-Ray Bursts, Baryon Acoustic Oscillations data, and Cosmic Microwave Background. In addition, the role of viscosity in the context of cosmic acceleration has been investigated in $f(Q)$ gravity \cite{R-1,R-2}. Moreover, the viability of $f(Q)$ cosmological models with respect to energy conditions, was verified by Mandal et al. \cite{Mandal/2020}. In such a work, the authors introduced the so-called embedding procedure, which enables one to include non-trivial contributions from the non-metricity function into the energy conditions. In a recent study, people have successfully used $f(Q)$ gravity in the Casimir wormholes \cite{Ghosh1} as well as the GUP-corrected Casimir wormhole \cite{Ghosh2}.
 The $f(Q)$ gravity was extended by Xu et al. \cite{Y.Xu et al} to the so-called $f(Q,T)$ gravity, where the $T$ stands for the trace of the energy-momentum tensor. Such a trace is responsible to add extra contributions from the quantum domain to classical gravity. Recently Arora et al. unveiled the viability of cosmological models for $f(Q,T)$ gravity with respect to energy conditions \cite{Simran/2021} and Tayde et al. \cite{Tayde/2023} used the Israel junction condition to see the stability of a thin-shell around the wormhole and potential caused by it.

 In this investigation, we apply the embedding procedure to derive traversable wormhole solutions in $f(Q,T)$ gravity with a radial dependent Bag parameter. The embedding procedure enables us to insert the nontrivial contributions coming from $f(Q,T)$ gravity into the effective equations for density and pressures. This methodology allows us to use the Raychaudhuri equations to derive the constraints over the $f(Q,T)$ families of traversable wormholes. In our studies, we analyze the features of the weak, the null, the dominant, and the strong energy conditions for wormholes in the presence of an anisotropic fluid.
 
  Along with our discussions, we carefully studied one family of $f(Q,T)$ gravity, whose wormhole solutions were derived using two different shape functions,  a $r_0 \gamma  \left(1-\frac{r_0}{r}\right)+r_0$ and $r_0\,\left(\frac{r}{r_0}\right)^{\lambda}$ ones.  We are going to present scenarios with the possibility of traversable wormholes, satisfying at least two sets of energy conditions in the presence of strange matter. We also derived the variation of the bag parameter with respect to the radial coordinate for the wormhole solutions. Moreover, we proved the stability of these wormholes through the equilibrium conditions coming from the generalized Tolman-Oppenheimer-Volkov equation. 

 The ideas presented in this work are summarized in the following nutshell. In sec. \ref{sec2}, we presented the generalities about the $f(Q,T)$ gravity. In \ref{sec3} we reviewed subjects on the derivation of the equation of motion for the MIT bag model. Then, in sec. \ref{sec4} we analyzed the energy conditions for the wormhole solutions in $f(Q,T)$ gravity, yielding traversable wormholes which obey at least two sets of energy conditions. In \ref{sec5} we presented the equilibrium conditions for these families of wormhole solutions. Our final remarks and perspectives for further works are described in sec.\ref{sec6}.
 
\section{Generalities in $f(Q,T)$ gravity}
\label{sec2}
We start the generalities on $f(Q,T)$ gravity by considering the following action 
\begin{equation}\label{1}
\mathcal{S}=\int\frac{1}{16\pi}\,f(Q,T)\sqrt{-g}\,d^4x+\int \mathcal{L}_m\,\sqrt{-g}\,d^4x\,,
\end{equation}
where $f(Q,T)$ is a function of non-metricity $Q$ and trace of the energy momentum tensor $T$, $g$ is the determinant of the metric $g_{\mu\nu}$, and $\mathcal{L}_m$ is the matter Lagrangian density. Such an action was introduced by Xu et al. in \cite{Y.Xu et al} where the reader can find more details on the generalities about the $f(Q,T)$ gravity. The non-metricity tensor is explicitly written as \cite{Jimenez}\\
\begin{equation}\label{2}
Q_{\lambda\mu\nu}=\bigtriangledown_{\lambda} g_{\mu\nu}\,.
\end{equation}
There the authors introduced the so-called $f(Q)$ gravity and present a detailed description of the generalities presented in this section.
Another relevant object of this theory is the non-metricity conjugate or superpotential, whose form is
\begin{equation}\label{4}
P^\alpha\;_{\mu\nu}=\frac{1}{4}\left[-Q^\alpha\;_{\mu\nu}+2Q_{(\mu}\;^\alpha\;_{\nu)}+Q^\alpha g_{\mu\nu}-\tilde{Q}^\alpha g_{\mu\nu}-\delta^\alpha_{(\mu}Q_{\nu)}\right]\,,
\end{equation}
where
\begin{equation}
\label{3}
Q_{\alpha}=Q_{\alpha}\;^{\mu}\;_{\mu}\,, \qquad \tilde{Q}_\alpha=Q^\mu\;_{\alpha\mu}.
\end{equation}
are the traces of the non-metricity tensor. From the previous definition one can derive the non-metricity scalar by taking the following contraction \cite{Jimenez}
\begin{eqnarray}
\label{5}
Q &=& -Q_{\alpha\mu\nu}\,P^{\alpha\mu\nu}\\
&=& -g^{\mu\nu}\left(L^\beta_{\,\,\,\alpha\mu}\,L^\alpha_{\,\,\,\nu\beta}-L^\beta_{\,\,\,\alpha\beta}\,L^\alpha_{\,\,\,\mu\nu}\right)\,,
\end{eqnarray}
where the disformation $L^\beta_{\,\,\,\mu\nu}$ is defined as
\begin{equation}
L^\beta_{\,\,\,\mu\nu}=\frac{1}{2}Q^\beta_{\,\,\,\mu\nu}-Q_{(\mu\,\,\,\,\,\,\nu)}^{\,\,\,\,\,\,\beta}.
\end{equation}

Now, by varying the action with respect to the metric tensor $g_{\mu\nu}$ we yield to the equations of motion for $f(Q,T)$ gravity, whose form is
\begin{equation}\label{7}
\frac{-2}{\sqrt{-g}}\bigtriangledown_\alpha\left(\sqrt{-g}\,f_Q\,P^\alpha\;_{\mu\nu}\right)-\frac{1}{2}g_{\mu\nu}f \\
+f_T \left(T_{\mu\nu} +\Theta_{\mu\nu}\right)\\
-f_Q\left(P_{\mu\alpha\beta}\,Q_\nu\;^{\alpha\beta}-2\,Q^
{\alpha\beta}\,\,_{\mu}\,P_{\alpha\beta\nu}\right)=8\pi T_{\mu\nu},
\end{equation}
here $f_Q=\frac{\partial f}{\partial Q}$ and $f_T=\frac{\partial f}{\partial T}$. Moreover,
\begin{equation}\label{6a}
\Theta_{\mu\nu}=g^{\alpha\beta}\frac{\delta T_{\alpha\beta}}{\delta g^{\mu\nu}}\,,
\end{equation}
and the energy-momentum tensor is given by
\begin{equation}\label{6}
T_{\mu\nu}=-\frac{2}{\sqrt{-g}}\frac{\delta\left(\sqrt{-g}\,\mathcal{L}_m\right)}{\delta g^{\mu\nu}}\,.
\end{equation}

\section{Equation of State for the MIT bag model}\label{sec3}

Several works have been connecting phenomenological astrophysical effects to the presence of strange quark matter, as was beautifully described in the work of Witten \cite{Witten}. There, it is highlighted that if one considers a star made of quark matter and ignores the contributions due to their masses, the relation between pressure and density is such that, $p=\frac{1}{3}(\rho - 4B)$, which is called MIT bag model, where the vacuum pressure $B$ on the Bag wall is responsible to stabilize the confinement of the quarks. The de-confinement phase transition depends on the temperature and the baryon number density of the system at high densities. This equation of state has been broadly applied in the literature, as we can see in \cite{Prasad,Aguirre,Burgio,Liu,Peng,Chakrabarty 1,Chakrabarty 2,Chakrabarty 3,N. Itoh}. In order to justify the form of the equation of state for MIT bag model, let us carefully revise the discussions presented by \cite{Deb/2022}. 

As it is known, the energy density and pressure of a system of particles can be determined from a general ensemble theory, and the physical features of such a system of particles may be derived from the following partition function
\begin{equation} \label{8}
\mathcal {Z}=\sum _{N_i, \epsilon } e^{ - \zeta (E_{N_i, \epsilon } - \sum _i N_i \mu _i) } \,,
\end{equation}
here $\zeta = \frac{1}{ k_B T}$, $N_i$ stands for the particle number, and $\mu_i$ is the chemical potential. We can realize that in general, the microscopic energy $E_{N_i, \epsilon }$ is a function of the particle number $N_i$, the masses of the particles, volume of the system $V$, and other quantum numbers $\epsilon$. Therefore, we may write $E_{N_i, \epsilon } = f(N_i, m_i, V, \epsilon )$. The pressure of the system is derived from the following relation
\begin{eqnarray}\label{9}
p_0 = -\Xi = \frac{1}{\zeta V} \ln \mathcal {Z}\,,
\end{eqnarray}
where $\Xi$, represents the thermodynamic potential which depend on chemical potential $\mu_i$, the mass of the particle $m_i$, and the temperature $T$. Furthermore, the statistical average of the energy density is such that
\begin{eqnarray}\label{11}
\bar{E} = - \frac{\partial }{\partial \zeta } \ln \mathcal {Z} + \sum _i \bar{N}_i \mu _i\,,
\end{eqnarray}
where the particle number $N_i$ has the form
\begin{eqnarray}\label{12}
\bar{N}_i = \frac{1}{\zeta } \left( \frac{\partial }{\partial \mu _i} \ln \mathcal {Z} \right) _{T, V, m_j} =-V \left( \frac{\partial \Xi }{\partial \mu _i} \right) _{T, m_j}\,.
\end{eqnarray}
We can also observe that the energy of the system is evaluated as
\begin{eqnarray}\label{13}
E_0 = \Xi + \sum _i n_i \mu _i\,,
\end{eqnarray}
and also that number density of particles is such that
\begin{equation}\label{14}
n_i = \frac{\bar{N}_i }{V} =- \left( \frac{\partial \Xi }{\partial \mu _i}\right) _{T, m_j} \,.
\end{equation}
In the case of the MIT bag model, the microscopic energy of the strange matter system is  
\begin{equation}\label{15}
E_{N_i, \epsilon }^{Bag}= E_{N_i, \epsilon } + B V\,,
\end{equation}
with the correspondent partition function
\begin{eqnarray}\label{16}
\mathcal {Z}^{Bag} =\mathcal {Z} e^{- \zeta B V}\,.
\end{eqnarray}
In this case, the number density of particles has the form
\begin{equation}\label{17}
n_i^{Bag} = \frac{\bar{N}_i}{V} =  - \left( \frac{\partial }{\partial \mu _i} (\Xi +B) \right) _{T, m_j, E_{N_i, \epsilon }, B }\,,
\end{equation}
where the bag pressure and energy are
\begin{equation}\label{18}
p^{Bag} = - (\Xi +B)\,; \qquad E^{Bag} = (\Xi +B) + \sum _i n_i \mu _i\,.
\end{equation}
So, by combining these previous equations with \eqref{9} and \eqref{13}, we find that
\begin{equation}
    p^{Bag}= p_0 - B\,; \qquad E^{Bag} = (E_0 +B)\,.
\end{equation}
As its known, for a relativistic fluid $p_0=E_0/3$. Consequently, by defining that $E^{Bag}=\rho$, we may write $p^{Bag}$ as
\begin{equation}
    p^{Bag} = \frac{1}{3}\,\left(\rho-4\,B\right)\,,
\end{equation}
proving the equation of state for the MIT bag model. In our next investigations, we are going to find wormhole families which obey the set of energy conditions in $f(Q,T)$ gravity, whose density and radial pressure obey the equation of state for the MIT bag model. We also intend to map the features of a bag parameter which depends on the radial coordinate. 

\section{Wormholes in $f(Q,T)$ gravity}
\label{sec4}
In order to approach the generalities on the wormhole solutions in $f(Q,T)$ gravity, let us consider the spherically symmetric and static wormhole metric in Schwarzschild coordinates $(t,\,r,\,\theta,\,\Phi)$ following the seminal work of Morris and Thorne \cite{Visser,Morris/1988}, whose specific form is
\begin{equation}\label{23}
ds^2=e^{2\phi(r)}dt^2-\left(1-\frac{b(r)}{r}\right)^{-1}dr^2-r^2\,d\theta^2-r^2\,\sin^2\theta\,d\Phi^2\,,
\end{equation}
where $\phi(r)$ and $b(r)$ denote the redshift function and the shape function, respectively. Here we are considering that the wormhole solutions for $f(Q,T)$ are compatible with the Birkhoff's Theorem, since we are writing the line element in the Schwarzschild form. Our conjecture on the viability of the Birkhoff's Theorem is based on the work of Meng and Wang \cite{Meng/2011}, and also on the recent review of Bahamond et al. \cite{Bahamonde/2023}. In their investigation, Meng and Wang proved the viability of the Birkhoff Theorem for teleparallel gravity \cite{Meng/2011}. Recently, Bahamonde et al. revisited the viability of the Birkhoff's Theorem for a generalized form of teleparallel gravity containing contributions of a boundary term and also coupled with scalar fields.  There they showed that the Birkhoff theorem has restrictions only if the scalars depend on $t$ or $r$ coordinates. Bahamonde et al. also unveiled in Eq. (5.49) of \cite{Bahamonde/2023}, a mapping transformation between the torsion and the non-metricity scalar. Such a mapping, together with the discussions on the Birkhoff's Theorem for teleparallel gravity, support our conjecture over the viability of such a theorem for solutions of $f(Q,T)$ gravity.

In our search for wormhole solutions we consider that the redshift and the shape functions obey the following constraints \cite{Visser,Morris/1988}:
\begin{itemize}
  \item[(1)] For $r>r_0$, i.e., out of throat, $1-\frac{b(r)}{r}>0$, and at the wormhole's throat i.e., $r=r_0$, $b(r)$ must satisfy the condition $b(r_0)=r_0$.
  \item[(2)] The shape function $b(r)$ has to fulfill the flaring-out requirement at the throat i.e., $b'(r_0)<1$.
  \item[(3)] For asymptotic flatness condition, the limit $\frac{b(r)}{r}\rightarrow 0$ as $r\rightarrow \infty$ is required.
  \item[(4)] Redshift function $\phi(r)$ should be finite everywhere.
\end{itemize}
In the present study, we assume the matter content of wormhole solutions is described by an anisotropic energy-momentum tensor which is given by \cite{Visser,Morris/1988}
\begin{equation}\label{24}
T_{\mu}^{\nu}=\left(\rho+p_t\right)u_{\mu}\,u^{\nu}-p_t\,\delta_{\mu}^{\nu}+\left(p_r-p_t\right)v_{\mu}\,v^{\nu},
\end{equation}
where, $\rho$ denotes the energy density, $u_{\mu}$ and $v_{\mu}$ are the four-velocity vector and unitary space-like vectors, respectively. Moreover, $p_r$ and $p_t$ are the radial and tangential pressures and both are functions of radial coordinate $r$. The anisotropic energy-momentum tensor was introduced by Letelier \cite{Letelier/1980} as a path to investigate a two-fluid model in plasma physics. Moreover, it has been applied in several distinct scenarios for modeling magnetized neutron stars \cite{Debabrata/2021}. Also, both four velocities satisfy the conditions $u_{\mu}u^{\nu}=-v_{\mu}v^{\nu}=1$, and from Eq. \eqref{24}, one can see that the trace of the energy-momentum tensor is such that $T=\rho-p_r-2p_t$.\\

In this investigation, we follow the procedures adopted by Moraes et al. \cite{Correa} in finding static wormhole solutions for $f(R,T)$. There the authors choose $\mathcal{L}_m=-P$, where $P=\frac{p_r+2\,p_t}{3}$ standing for the total pressure. Such Lagrangian changes Eq. \eqref{6a} to
\begin{equation}
    \Theta_{\mu\nu}=-g_{\mu\nu}\,P-2\,T_{\mu\nu}\,.
\end{equation}
Besides, the non-metricity scalar $Q$ for the metric \eqref{23} is explicit written as \cite{Tayde}
\begin{equation}\label{25}
Q=-\frac{b}{r^2}\left[\frac{rb^{'}-b}{r(r-b)}+2\phi^{'}\right].
\end{equation}\\
Now, by inserting the Eqs. \eqref{23}, \eqref{24} and \eqref{25} into the equation of motion \eqref{7}, we find
\begin{equation}\label{24a1}
\frac{2 (r-b)}{(2 r-b) f_Q}\left[\rho-\frac{(r-b)}{8 \pi  r^3} \left(\frac{b r f_{\text{QQ}} Q'}{r-b}+b f_Q \left(\frac{ r \phi '+1}{r-b}-\frac{2 r-b}{2 (r-b)^2}\right)+\frac{f r^3}{2 (r-b)}\right)+\frac{f_T (P+\rho )}{8 \pi }\right]=\frac{b'}{8 \pi  r^2},
\end{equation}
\begin{multline}\label{24a2}
\frac{2 b}{f r^3}\left[p_r +\frac{(r-b)}{16 \pi r^3} \left(f_Q \left(\frac{b \left(\frac{r b'-b}{r-b}+2 r \phi '+2\right)}{r-b}-4 r \phi '\right)+\frac{2 b r f_{\text{QQ}} Q'}{r-b}\right)+\frac{fr^3 (r-b)\phi '}{8\pi b  r^2}-\frac{f_T \left(P-p_r\right)}{8 \pi }\right]\\
=\frac{1}{8 \pi}\left[2\left(1-\frac{b}{r}\right)\frac{\phi '}{r}-\frac{b}{r^3}\right],
\end{multline}
\begin{multline}\label{24a3}
\frac{1}{f_Q \left(\frac{r}{r-b}+r \phi '\right)}\left[p_t +\frac{(r-b)}{32 \pi r^2}\left(f_Q \left(\frac{4 (2 b-r) \phi '}{r-b}-4 r \left(\phi '\right)^2-4 r \phi ''\right)+\frac{2 f r^2}{r-b}-4 r f_{\text{QQ}} Q' \phi '\right) \right.\\ \left.
+\frac{(r-b)}{8\pi r}\left(\phi '' +{\phi '}^2-\frac{(rb'-b)\phi '}{2r(r-b)}+\frac{\phi '}{r}\right)f_Q \left(\frac{r}{r-b}+r \phi '\right)-\frac{f_T \left(P-p_t\right)}{8 \pi }\right]\\
=\frac{1}{8\pi}\left(1-\frac{b}{r}\right)\left[\phi '' +{\phi '}^2-\frac{(rb'-b)\phi '}{2r(r-b)}-\frac{rb'-b}{2r^2 (r-b)}+\frac{\phi '}{r}\right]\,,
\end{multline}\\
which are the non-zero components of the field equations for $f(Q,T)$ gravity \cite{Tayde}.
In order to properly determine the constraints over the pressures and density for our wormhole solutions, let us consider the Morris-Throne field equations for traversable wormholes, whose forms for GR are
\begin{equation}\label{24a4}
\frac{b'}{8 \pi  r^2}= \tilde{\rho},
\end{equation}
\begin{equation}\label{24a5}
\frac{1}{8 \pi}\left[2\left(1-\frac{b}{r}\right)\frac{\phi '}{r}-\frac{b}{r^3}\right] = \tilde{p_r},
\end{equation}
\begin{equation}\label{24a6}
\frac{1}{8\pi}\left(1-\frac{b}{r}\right)\left[\phi '' +{\phi '}^2-\frac{(rb'-b)\phi '}{2r(r-b)}-\frac{rb'-b}{2r^2 (r-b)}+\frac{\phi '}{r}\right] = \tilde{p_t}.
\end{equation}
where $\tilde{\rho}$,$\tilde{p_r}$ and $\tilde{p_t}$ are corresponding energy density, radial pressure and tangential pressure, respectively. So, by comparing Eqs. \eqref{24a1}-\eqref{24a3} with the Eqs. \eqref{24a4}-\eqref{24a6}, we get 
\begin{equation}\label{24a7}
\tilde{\rho}=\frac{2 (r-b)}{(2 r-b) f_Q}\left[\rho-\frac{1}{8 \pi  r^2}\left(1-\frac{b}{r}\right) \left(\frac{b r f_{\text{QQ}} Q'}{r-b}+b f_Q \left(\frac{ r \phi '+1}{r-b}-\frac{2 r-b}{2 (r-b)^2}\right)+\frac{f r^3}{2 (r-b)}\right)+\frac{f_T (P+\rho )}{8 \pi }\right],
\end{equation}
\begin{equation}\label{24a8}
\tilde{p_r}=\frac{2 b}{f r^3}\left[p_r +\frac{1}{16 \pi r^2}\left(1-\frac{b}{r}\right) \left(f_Q \left(\frac{b \left(\frac{r b'-b}{r-b}+2 r \phi '+2\right)}{r-b}-4 r \phi '\right)+\frac{2 b r f_{\text{QQ}} Q'}{r-b}\right)+\frac{fr^3 (r-b)\phi '}{8\pi b  r^2}-\frac{f_T \left(P-p_r\right)}{8 \pi }\right],
\end{equation}
\begin{multline}\label{24a9}
\tilde{p_t}=\frac{1}{f_Q \left(\frac{r}{r-b}+r \phi '\right)}\left[p_t +\frac{1}{32 \pi r}\left(1-\frac{b}{r}\right) \left(f_Q \left(\frac{4 (2 b-r) \phi '}{r-b}-4 r \left(\phi '\right)^2-4 r \phi ''\right)+\frac{2 f r^2}{r-b}-4 r f_{\text{QQ}} Q' \phi '\right) \right.\\ \left.
+\frac{1}{8\pi}\left(1-\frac{b}{r}\right)\left(\phi '' +{\phi '}^2-\frac{(rb'-b)\phi '}{2r(r-b)}+\frac{\phi '}{r}\right)f_Q \left(\frac{r}{r-b}+r \phi '\right)-\frac{f_T \left(P-p_t\right)}{8 \pi }\right].
\end{multline}\\
The last equations unveiled that the non-trivial contributions from $f(Q,T)$ gravity were embedded into $\tilde{\rho}$,$\tilde{p_r}$ and $\tilde{p_t}$, which we are going to consider as the effective density, radial and tangential pressures. This embedding procedure was introduced by Sanjay et al. \cite{Mandal/2020} in determining energy conditions to $f(Q)$ gravity. It was also applied by Hassan et al. \cite{Hassan/2022} to constraint energy conditions for traversable wormholes in $f(Q)$ gravity. 

Then, from Eqs. \eqref{24a1}-\eqref{24a3}, one can find the following field equations for $f(Q,T)$ gravity \cite{Tayde}:
\begin{equation}\label{24a10}
8 \pi  \rho =\frac{(r-b)}{2 r^3} \left[f_Q \left(\frac{(2 r-b) \left(r b'-b\right)}{(r-b)^2}+\frac{b \left(2 r \phi '+2\right)}{r-b}\right)+\frac{2 b r f_{\text{QQ}} Q'}{r-b}+\frac{f r^3}{r-b}-\frac{2r^3 f_T (P+\rho )}{(r-b)}\right],
\end{equation}
\begin{equation}\label{24a11}
8 \pi  p_r=-\frac{(r-b)}{2 r^3} \left[f_Q \left(\frac{b }{r-b}\left(\frac{r b'-b}{r-b}+2 r \phi '+2\right)-4 r \phi '\right)+\frac{2 b r f_{\text{QQ}} Q'}{r-b}+\frac{f r^3}{r-b}-\frac{2r^3 f_T \left(P-p_r\right)}{(r-b)}\right],
\end{equation}
\begin{equation}\label{24a12}
8 \pi  p_t=-\frac{(r-b)}{4 r^2} \left[f_Q \left(\frac{\left(r b'-b\right) \left(\frac{2 r}{r-b}+2 r \phi '\right)}{r (r-b)}+\frac{4 (2 b-r) \phi '}{r-b}-4 r \left(\phi '\right)^2-4 r \phi ''\right)-4 r f_{\text{QQ}} Q' \phi '+\frac{2 f r^2}{r-b}-\frac{4r^2 f_T \left(P-p_t\right)}{(r-b)}\right].
\end{equation}\\

As pointed by  Morris and Thorne, the redshift function $\phi(r)$ must be finite everywhere in order to avoid horizons in the traversable wormholes \cite{Morris/1988}. Therefore, one simple form of $\phi(r)$ which is finite and also allows one to find analytic constraints over the energy conditions is $\phi(r) = constant$. This form of the redshift function implies that the effective density and pressures are given by
\begin{equation}\label{24a13}
\tilde{\rho}=\frac{2 (r-b) }{(2 r-b) f_Q}\left(\rho-\frac{\left(1-\frac{b}{r}\right) \left(\frac{b r f_{\text{QQ}} Q'}{r-b}+\frac{b f_Q}{r-b}-\frac{b (2 r-b) f_Q}{2 (r-b)^2}+\frac{f r^3}{2 (r-b)}\right)}{8 \pi  r^2}+\frac{f_T (P+\rho )}{8 \pi }\right)\,,
\end{equation}
\begin{equation}\label{24a14}
\tilde{p_r}=\frac{2 b }{f r^3}\left(p_r- \frac{f_T \left(P-p_r\right)}{8 \pi }+\frac{\left(1-\frac{b}{r}\right) \left(\frac{b f_Q \left(\frac{r b'-b}{r-b}+2\right)}{r-b}+\frac{2 b r f_{\text{QQ}} Q'}{r-b}\right)}{16 \pi  r^2}\right)\,,
\end{equation}
\begin{equation}\label{24a15}
\tilde{p_t}=\frac{(r-b)}{r f_Q}\left(p_t-\frac{f_T \left(P-p_t\right)}{8 \pi }+ \frac{f r \left(1-\frac{b}{r}\right)}{16 \pi  (r-b)}\right)\,.
\end{equation}

Let us dedicate a few lines to standard energy conditions derived using the Raychaudhuri equations. The Raychaudhuri equations enable us to describe the action of congruence and attractiveness of the gravity for timelike, spacelike, or lightlike curves. By following the prescriptions adopted by Arora et al. \cite{Simran/2021}, the Raychaudhuri equations impose the following constraints over wormhole's density and pressures:\\
$\bullet$ Weak energy conditions (WEC) if $\tilde{\rho}\geq0$, $\tilde{\rho}+\tilde{p_j}\geq0$, $\forall j$.\\
$\bullet$ Null energy condition (NEC) if $\tilde{\rho}+\tilde{p_j}\geq0$, $\forall j$.\\ 
$\bullet$ Dominant energy conditions (DEC) if $\tilde{\rho}\geq0$, $\tilde{\rho} \pm \tilde{p_j}\geq0$, $\forall j$.\\
$\bullet$ Strong energy conditions (SEC) if $\tilde{\rho}+\tilde{p_j}\geq0$, $\tilde{\rho}+\sum_j\tilde{p_j}\geq0$, $\forall j$.\\
where $j=r,\,t$.\\
So, taking the effective density and pressures into the account from eqs. \eqref{24a13}-\eqref{24a15}, we get
\begin{multline}\
\tilde{\rho} + \tilde{p_r}= \frac{b (b-2 r) f_Q^2 \left(b (3 b-2 r)-b r b'\right)-f r^4 (b-r)^2 \left(2 b f_{\text{QQ}} Q'+r^2 \left(-2 f_T (P+\rho )+f+16 \pi  p_r\right)\right)}{8 \pi  f r^6 (b-2 r) (b-r) f_Q}\\
+\frac{2 (r-b) \left(\rho+p_r \right)}{(2 r-b) f_Q}+\frac{b r (b-r) f_Q \left(2 (b-2 r) \left(b f_{\text{QQ}} Q'+r^2 \left(f_T \left(p_r-P\right)+8 \pi  p_r\right)\right)-b f r^2\right)}{8 \pi  f r^6 (b-2 r) (b-r) f_Q}\,,
\end{multline}
\begin{equation}
\tilde{\rho} + \tilde{p_t}=\frac{2 (r-b) \left(\rho+p_t \right)}{(2 r-b) f_Q}-\frac{2 b^2 f_Q+r (r-b) \left(2 r f_T \left(-b p_t+b P+2 r p_t+2 \rho  r\right)-b \left(4 f_{\text{QQ}} Q'+f r+16 \pi  r p_t\right)\right)}{16 \pi  r^3 (b-2 r) f_Q}\,,
\end{equation}
\begin{multline}
\tilde{\rho} - \tilde{p_r}=-\frac{b (b-2 r) f_Q^2 \left(b (3 b-2 r)-b r b'\right)+f r^4 (b-r)^2 \left(2 b f_{\text{QQ}} Q'+r^2 \left(-2 f_T (P+\rho )+f-16 \pi  p_r\right)\right)}{8 \pi  f r^6 (b-2 r) (b-r) f_Q}\\
+\frac{2 (r-b) \left(\rho -p_r\right)}{(2 r-b) f_Q}-\frac{b r (b-r) f_Q \left(2 (b-2 r) \left(b f_{\text{QQ}} Q'+r^2 \left(f_T \left(p_r-P\right)+8 \pi  p_r\right)\right)+b f r^2\right)}{8 \pi  f r^6 (b-2 r) (b-r) f_Q}\,,
\end{multline}
\begin{equation}
\tilde{\rho} - \tilde{p_t}=\frac{2 (r-b) \left(\rho -p_t\right)}{(2 r-b) f_Q}-\frac{2 b^2 f_Q+r (r-b) \left(2 \left(r f_T \left((b-2 r) p_t-b P+4 P r+2 \rho  r\right)-2 b f_{\text{QQ}} Q'+8 \pi  b r p_t\right)+f r (b-4 r)\right)}{16 \pi  r^3 (b-2 r) f_Q}\,,
\end{equation}
\begin{multline}
\hspace{-0.5cm}\tilde{\rho} + \tilde{p_r} + 2\tilde{p_t}=\frac{b^2 f_Q \left(-r b'+3 b-2 r\right)}{8 \pi r^6\,f\, (b-r)}+\frac{r^4 (b-r) \left(2 r f_T \left(-(b-2 r) p_t+b P-P r+\rho  r\right)-2 b f_{\text{QQ}} Q'+f r (r-b)-16 \pi  r \left(b p_t+r p_r\right)\right)}{8 \pi r^6\,(b-2 r) f_Q}\\
+\frac{2 (r-b) \left(p_r+2 p_t+\rho \right)}{(2 r-b) f_Q}+\frac{b\, r }{8 \pi  r^6}\left(\frac{2 \left(b f_{\text{QQ}} Q'+r^2 \left(f_T \left(p_r-P\right)+8 \pi  p_r\right)\right)}{f}-\frac{b r^2}{b-2 r}\right)\,.
\end{multline}
Therefore, the energy conditions for wormhole at $f(Q,T)$ gravity from previous equations are

$\bullet$ \textbf{WEC \,:} \quad $\tilde{\rho}\geq0$ \quad$\Rightarrow$\quad $\rho \geq0$ \,\,\,\,\,\,\, with \,\,\,\,\,\,\, $\frac{(2 r-b) f_Q}{r-b} > 0$ , \,\,\, $\frac{2 (r-b) \left(\frac{\left(1-\frac{b}{r}\right) \left(\frac{b r f_{\text{QQ}} Q'}{r-b}+\frac{b f_Q}{r-b}-\frac{b (2 r-b) f_Q}{2 (r-b)^2}+\frac{f r^3}{2 (r-b)}\right)}{8 \pi  r^2}-\frac{f_T (P+\rho )}{8 \pi }\right)}{(2 r-b) f_Q} \leq0$\,,\\

$\bullet$ \textbf{NEC \,:} \quad $\tilde{\rho}+\tilde{p_r}\geq0$ \quad$\Rightarrow$\quad $\rho + p_r \geq0$ \,\,\,\,\,\,\, with \,\,\,\,\,\,\, $\frac{(2 r-b) f_Q}{r-b} > 0$ , \,\,\, $\frac{b r (b-r) f_Q \left(2 (b-2 r) \left(b f_{\text{QQ}} Q'+r^2 \left(f_T \left(p_r-P\right)+8 \pi  p_r\right)\right)-b f r^2\right)}{8 \pi  f r^6 (b-2 r) (b-r) f_Q}\\
+\frac{b (b-2 r) f_Q^2 \left(b (3 b-2 r)-b r b'\right)-f r^4 (b-r)^2 \left(2 b f_{\text{QQ}} Q'+r^2 \left(-2 f_T (P+\rho )+f+16 \pi  p_r\right)\right)}{8 \pi  f r^6 (b-2 r) (b-r) f_Q} \geq0$ \quad and \\

$\tilde{\rho}+\tilde{p_t}\geq0$ \quad$\Rightarrow$\quad $\rho + p_t \geq0$ \,\,\,\,\, with \,\,\,\,\, $\frac{(2 r-b) f_Q}{r-b} > 0$ , \,\,\, $\frac{2 b^2 f_Q+r (r-b) \left(2 r f_T \left(-b p_t+b P+2 r p_t+2 \rho  r\right)-b \left(4 f_{\text{QQ}} Q'+f r+16 \pi  r p_t\right)\right)}{16 \pi  r^3 (b-2 r) f_Q} \leq0$\,,\\

$\bullet$ \textbf{DEC \,:} \quad $\tilde{\rho}-\tilde{p_r}\geq0$ \quad$\Rightarrow$\quad $\rho - p_r \geq0$ \,\,\,\,\,\,\, with \,\,\,\,\,\,\, $\frac{(2 r-b) f_Q}{r-b} > 0$ , \,\,\, $\frac{b r (b-r) f_Q \left(2 (b-2 r) \left(b f_{\text{QQ}} Q'+r^2 \left(f_T \left(p_r-P\right)+8 \pi  p_r\right)\right)+b f r^2\right)}{8 \pi  f r^6 (b-2 r) (b-r) f_Q}\\
+\frac{b (b-2 r) f_Q^2 \left(b (3 b-2 r)-b r b'\right)+f r^4 (b-r)^2 \left(2 b f_{\text{QQ}} Q'+r^2 \left(-2 f_T (P+\rho )+f-16 \pi  p_r\right)\right)}{8 \pi  f r^6 (b-2 r) (b-r) f_Q} \leq0$ \quad and \\

$\tilde{\rho}-\tilde{p_t}\geq0$ \quad$\Rightarrow$\quad $\rho - p_t \geq0$ \,\,\,\,\, with \,\,\,\,\, $\frac{(2 r-b) f_Q}{r-b} > 0$ ,\\
\qquad $\frac{2 b^2 f_Q+r (r-b) \left(2 \left(r f_T \left((b-2 r) p_t-b P+4 P r+2 \rho  r\right)-2 b f_{\text{QQ}} Q'+8 \pi  b r p_t\right)+f r (b-4 r)\right)}{16 \pi  r^3 (b-2 r) f_Q} \leq0$\,,\\

$\bullet$ \textbf{SEC \,:} \quad $\tilde{\rho}+\tilde{p_r} + 2\tilde{p_t} \geq0$ \quad$\Rightarrow$\quad $\rho + p_r + 2 p_t \geq0$ \,\,\,\,\,\,\, with \,\,\,\,\,\,\, $\frac{(2 r-b) f_Q}{r-b} > 0$ ,\\
$\frac{b r \left(\frac{2 \left(b f_{\text{QQ}} Q'+r^2 \left(f_T \left(p_r-P\right)+8 \pi  p_r\right)\right)}{f}-\frac{b r^2}{b-2 r}\right)}{8 \pi  r^6}+\frac{\frac{b^2 f_Q \left(-r b'+3 b-2 r\right)}{f (b-r)}+\frac{r^4 (b-r) \left(2 r f_T \left(-(b-2 r) p_t+b P-P r+\rho  r\right)-2 b f_{\text{QQ}} Q'+f r (r-b)-16 \pi  r \left(b p_t+r p_r\right)\right)}{(b-2 r) f_Q}}{8 \pi  r^6} \geq0$\,.\\

As its known, all the energy conditions previously presented are not obeyed in standard GR for traversable wormhole solutions. Especially the NEC violation indicates the presence of exotic matter at the throat of the wormhole, moreover, the energy density also needs to be positive for a realistic matter source of the wormhole solutions. \\
In order to test the viability of traversable wormhole solutions which also obey the MIT bag EoS, let us work with the following  $f(Q,T)$ function \cite{Y.Xu et al}
\begin{equation}
\label{2c}
f(Q,T)=\alpha\,Q+\beta\,T.
\end{equation}
where $\alpha$ and $\beta$ are free parameters.\\
Using the above linear functional form of $f(Q,T)$ and assuming the redshift function $\phi(r)=constant$ in the field equations \eqref{24a10}-\eqref{24a12},  we obtain
 \begin{equation}\label{29}
 \rho =\frac{\alpha  (12 \pi -\beta ) b'}{3 (4 \pi -\beta ) (\beta +8 \pi ) r^2}\,,
 \end{equation}
 \begin{equation}\label{30}
 p_r=-\frac{\alpha  \left(2 \beta  r b'-3 \beta  b+12 \pi  b\right)}{3 (4 \pi -\beta ) (\beta +8 \pi ) r^3}\,,
 \end{equation}
 \begin{equation}\label{31}
 p_t=-\frac{\alpha  \left((\beta +12 \pi ) r b'+3 b (\beta -4 \pi )\right)}{6 (4 \pi -\beta ) (\beta +8 \pi ) r^3}\,.
 \end{equation}
 
\subsubsection{\textbf{Wormhole solutions with different shape functions}}
In this subsection we consider a constant redshift function and that the radial pressure obeys a generalized version of the equation of state for the MIT bag model, whose form is
\begin{equation}\label{32}
p_r = \omega (\rho - 4B)\,,
\end{equation}
 where $\omega$ is EoS parameter and $B$ is the bag parameter. We can observe that this equation recovers the standard MIT bag model if $\omega=1/3$. Consequently, by taking equations \eqref{29} and \eqref{30} into eq. \eqref{32}, we obtain
 \begin{equation}\label{32_5}
B=\frac{\alpha  \left(-\beta  r \omega  b'+2 \beta  r b'+12 \pi  r \omega  b'-3 \beta  b+12 \pi  b\right)}{12 (4 \pi -\beta ) (\beta +8 \pi ) r^3 \omega }\,,
 \end{equation}
as a bag parameter whose form depends on the shape function $b(r)$. From the previous equation, we realize that the bag parameter indicates different energy domains for stabilizing the strange quark matter in different regions of the wormhole.  \\

\begin{figure}[H]
\centering
\includegraphics[width=7.5cm,height=5cm]{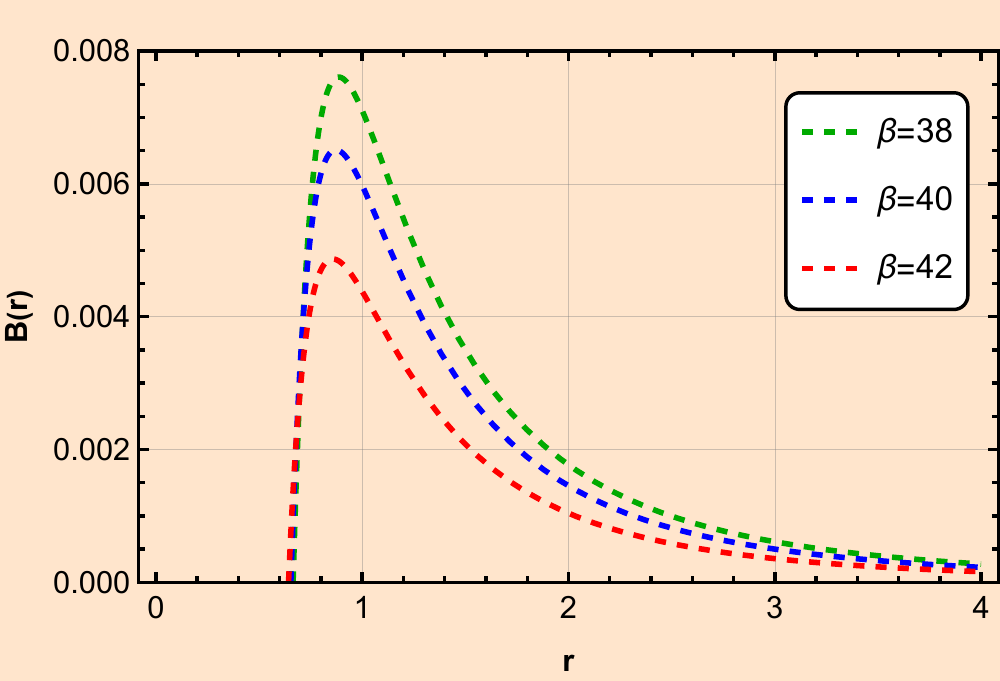}
\caption{The behavior of Bag parameter $B$ for (a) $\beta=38$ with $\omega =0.28$,\,\,\,\,(b) $\beta=40$ with $\omega =0.333$\,\,\, and (c) $\beta=42$ with $\omega =0.455$. Also, we consider $\alpha=1,\, \gamma=0.5,\,r_0=1$.}
\label{Fig1}
\end{figure}

\begin{figure}[H]
\centering
\includegraphics[width=7.5cm,height=5cm]{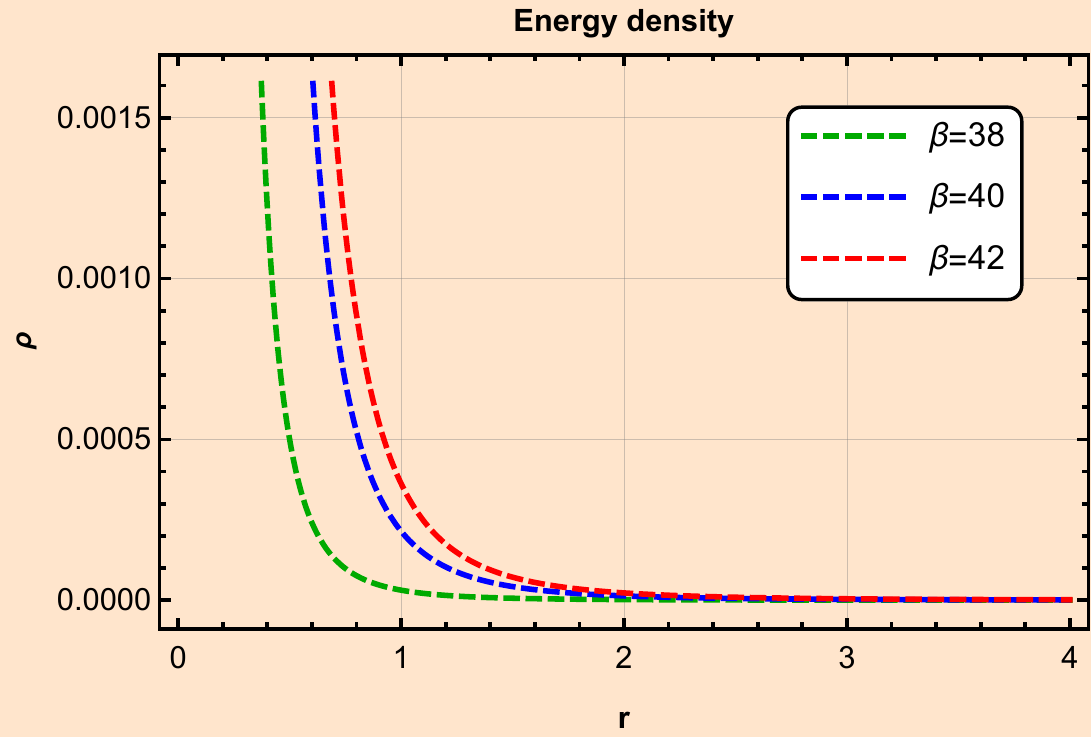}
\centering
\caption{The figure shows the behavior of $\rho$ with respect to $r$ for different valus of $\beta$. We consider $\alpha=1,\, \gamma=0.5,\,r_0=1$.}
\label{fig2}
\end{figure}

\begin{figure}[H]
\centering
\includegraphics[width=14.5cm,height=5cm]{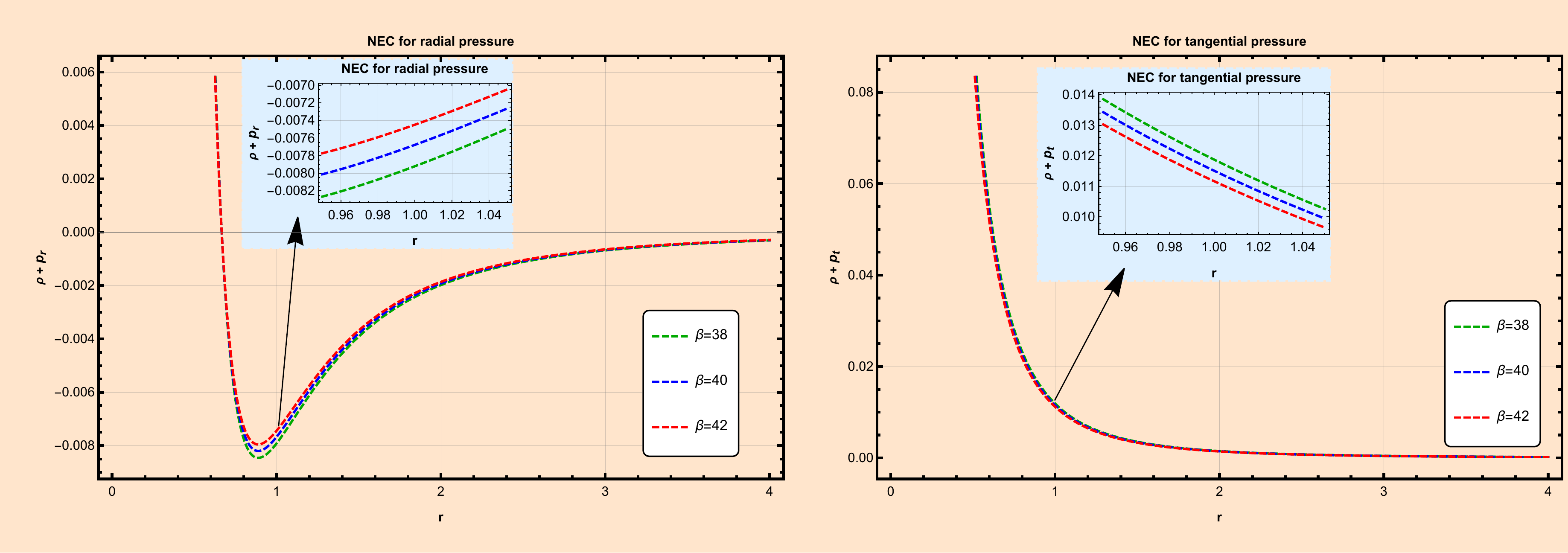}
\caption{Profile shows the behavior of NEC with respect to $r$ for different values of $\beta$. We consider $\alpha=1,\, \gamma=0.5,\,r_0=1$.}
\label{fig3}
\end{figure}

\begin{figure}[H]
\centering
\includegraphics[width=14.5cm,height=5cm]{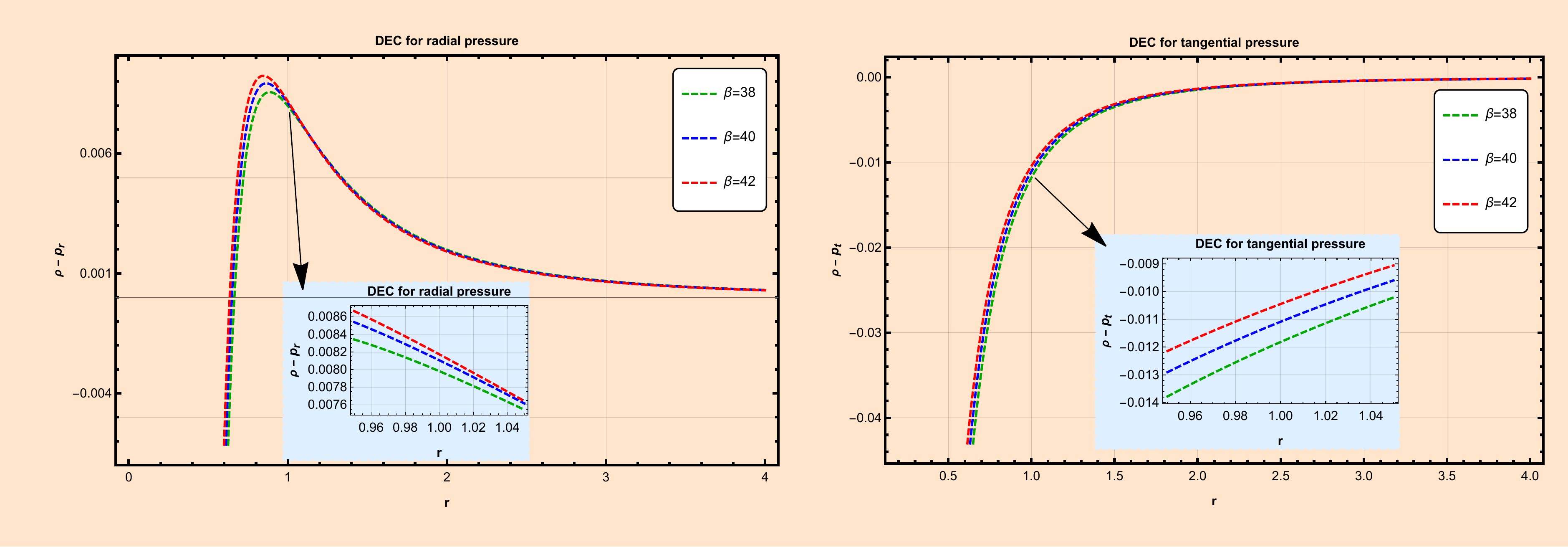}
\caption{Profile shows the behavior of DEC with respect to $r$ for different values of $\beta$. We consider $\alpha=1,\, \gamma=0.5,\,r_0=1$.}
\label{fig4}
\end{figure}

\begin{figure}[H]
\centering
\includegraphics[width=6.5cm,height=4cm]{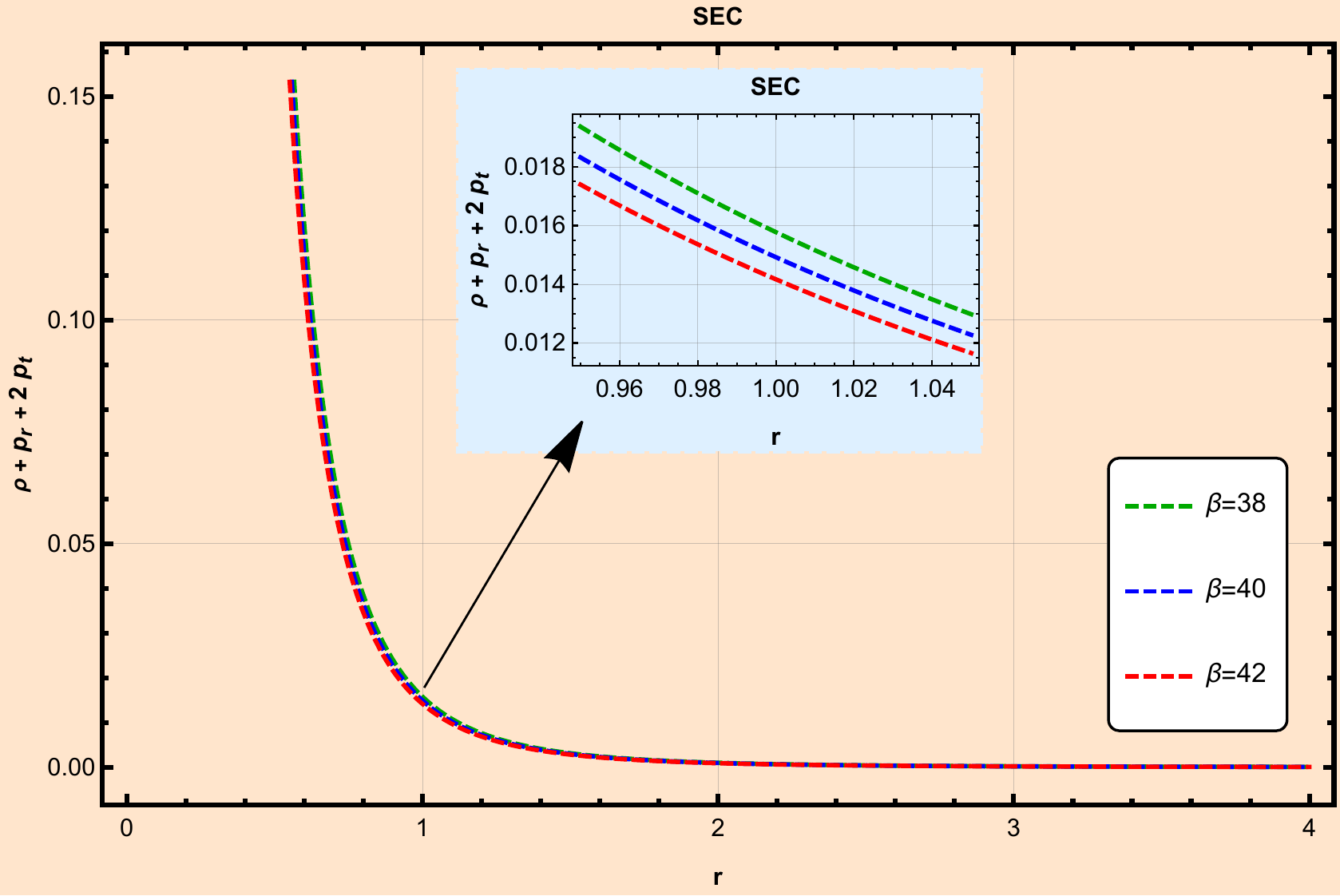}
\centering
\caption{Profile shows the behavior of SEC with respect to $r$ for different values of $\beta$. We consider $\alpha=1,\, \gamma=0.5,\,r_0=1$.}
\label{fig5}
\end{figure}

 \textbf{Shape Function-1 :} As a first example, let us consider a shape function introduced by Harko et al. \cite{T. Harko} on the study of electromagnetic signatures of the accretion disks of wormholes, whose form is 
 \begin{equation}\label{33}
 b(r)= r_0 \gamma  \left(1-\frac{r_0}{r}\right)+r_0
 \end{equation}
 where $\gamma<1$ in order to satisfy  flaring out condition $\left(b'(r)<1\right)$, the throat condition $\left(b(r)-r<0\right)$, and the asymptotically flatness condition $\left(\frac{b(r)}{r}\rightarrow 0 \,\,; r\rightarrow \infty\right)$. From Eqs. \eqref{32_5} and \eqref{33} we determine the following bag parameter
 \begin{equation}\label{34}
 B=\frac{r_0 \alpha  (12 \pi  (r_0 \gamma  (\omega -1)+(\gamma +1) r)-\beta  (r_0 \gamma  (\omega -5)+3 (\gamma +1) r))}{12 (4 \pi -\beta ) (\beta +8 \pi ) r^4 \omega }\,,
 \end{equation}
 which is depicted in details in Fig \ref{Fig1}. There we can observe that $B$ approaches to zero inside of the wormhole throat $\left(r=r_0\right)$, it also goes to constant values for big values of $r$. Moreover, as $\beta$ decreases, we have smaller maximum values for the bag parameter, resulting in smaller binding energy for the matter. This behavior of $B(r)$ was observed by Deb et al.  \cite{Deb/2022} for $f(R,T)$ gravity with a hybrid exponential shape function.  It is relevant to say that $\omega=0.455$ and $\omega=0.333$ here chosen, were used by Harko and Cheng \cite{Harko/2002} in their investigation on equations of state for strange stars and  $\omega=0.28$ is connected with a constraint over the mass of strange quarks as pointed by Deb et al. \cite{Deb/2022}.
Now,  by considering the Eqs. \eqref{2c} and \eqref{33}, we obtain the following set of energy conditions\\
$\bullet$ \textbf{NEC :} $\rho+p_r=-\frac{r_0 \alpha  (-2 r_0 \gamma +\gamma  r+r)}{(\beta +8 \pi ) r^4}$ and $\rho+p_t=\frac{r_0 \alpha  (\gamma +1)}{2 (\beta +8 \pi ) r^3}$,\\
$\bullet$ \textbf{DEC :} $\rho-p_r=\frac{r_0 \alpha  (4 r_0 \beta  \gamma -3 \beta  (\gamma +1) r+12 \pi  (\gamma +1) r)}{3 (4 \pi -\beta ) (\beta +8 \pi ) r^4}$ and $\rho-p_t=\frac{r_0 \alpha  (-4 r_0 \beta  \gamma -12 \pi  (-4 r_0 \gamma +\gamma  r+r)+3 \beta  (\gamma +1) r)}{6 (4 \pi -\beta ) (\beta +8 \pi ) r^4}$,\\
$\bullet$ \textbf{SEC :} $\rho+p_r+2p_t=-\frac{4 r_0^2 \alpha  \beta  \gamma }{3 (4 \pi -\beta ) (\beta +8 \pi ) r^4}$\,,\\
whose features are presented in Figs. \ref{fig2} - \ref{fig5}. There we can see that the energy density is positive over the entire variation of $r$, satisfying WEC. We also verify that NEC is violated by radial pressure and satisfied by tangential pressure, and SEC is satisfied. Moreover, DEC is violated inside of the wormholes throats for both radial and tangential pressures. The violation of DEC in this region is consistent with the presence of exotic matter inside of the wormholes throats, and they are compatible with the features presented on wormholes in $f(R,T)$ gravity, as we can see in \cite{Elizalde/2018}.  Besides, this partial violation of NEC can be justified by the behavior of Eq. \eqref{32} for positive values of $B$ and $\rho$. Once $B>\rho$ inside of the wormholes throat, then $p_r$ is going to be negative in this region, resulting in the partial violation of NEC. Therefore, we can connect such a violation with the presence of strange matter at the inner region of the wormhole.

It is relevant to point that these constraints over the energy conditions are also observed for $\tilde{\rho}$, $\tilde{p}_r$, and $\tilde{p}_t$. The violations of NEC and DEC are compatible with the presence of strange matter inside the wormholes throats, which may be formed by strange quarks. The fact that both WEC and SEC are satisfied indicates a scenario of stable traversable families of wormholes. 



\begin{figure}[H]
\centering
\includegraphics[width=7.5cm,height=5cm]{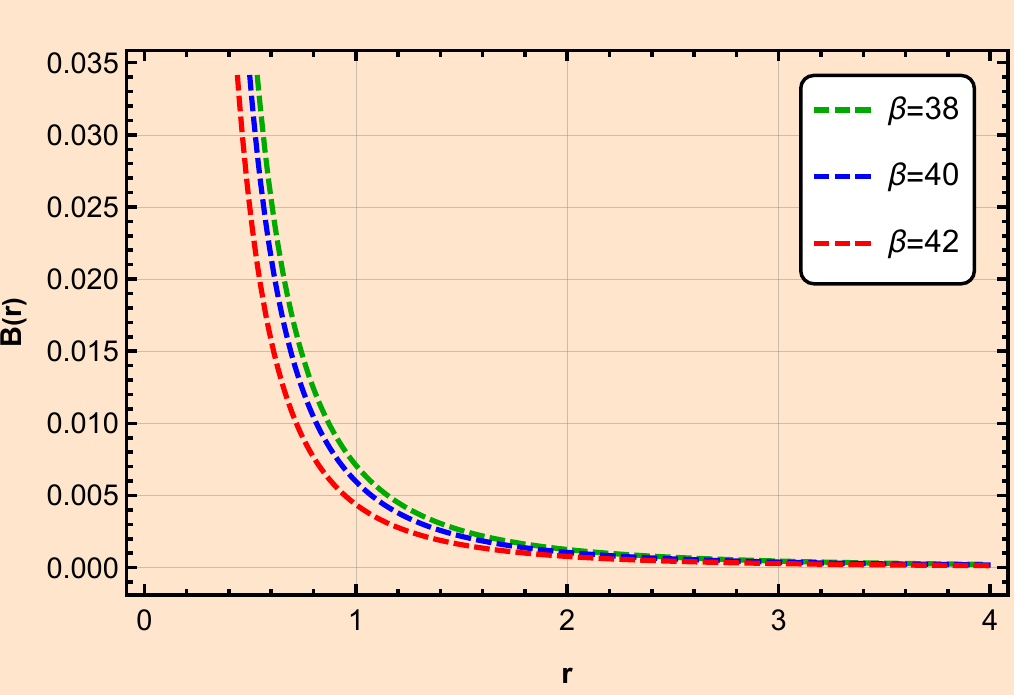}
\caption{The behavior of Bag parameter $B$ for (a) $\beta=38$ with $\omega =0.28$,\,\,\,\,(b) $\beta=40$ with $\omega =0.333$\,\,\, and (c) $\beta=42$ with $\omega =0.455$. Also, we consider $\alpha=1,\, r_0=1$.}
\label{fig6}
\end{figure}

\textbf{Shape Function-2 :}  We consider a shape function discussed in the seminal paper of Lobo et al. \cite{Lobo/2009}, where the authors investigated the viability of wormhole solutions for different $f(R)$ theories of gravity. This shape function is given by 
\begin{equation}\label{35}
b(r) = r_0\,\left(\frac{r}{r_0}\right)^{\lambda}\,.
\end{equation}
 where $\lambda <1$ in order to satisfy the flaring out condition $\left(b'(r)<1\right)$, throat condition $\left(b(r)-r<0\right)$ and asymptotically flatness condition $\left(\frac{b(r)}{r}\rightarrow 0 \,\,, r\rightarrow \infty\right)$. In order to find the energy constraints let us work with  $\lambda=\frac{1}{2}$, yielding to the shape function
 \begin{equation}\label{35(a)}
   b(r) = r_0 \sqrt{\frac{r}{r_0}}\,.
 \end{equation}
 With these ingredients in hands we can determine the following form for the bag parameter
 \begin{equation}\label{36}
 B=\frac{\alpha  (12 \pi  (\omega +2)-\beta  (\omega +4))}{24 (4 \pi -\beta ) (\beta +8 \pi ) r^2 \omega  \sqrt{\frac{r}{r_0}}}\,,
 \end{equation}
 whose features are presented in Fig. \ref{fig6}. There we observe that $B(r)$ is always positive for the different values of $\beta$ and $\omega$ here chosen. We also realize that $B(r)$ increases near to the wormhole throat, raising the binding energy to keep the strange matter stable. Moreover, the bag parameter goes to a constant value for big values of $r$. These features also corroborates with the investigations of Deb et al. in \cite{Deb/2022}.

\begin{figure}[H]
\centering
\includegraphics[width=7.5cm,height=5cm]{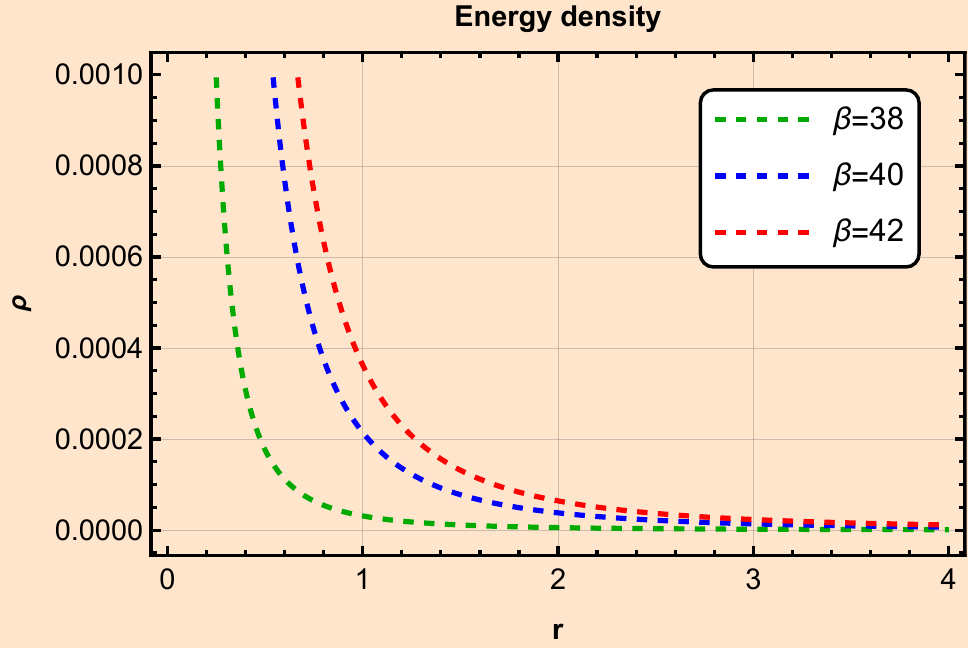}
\centering
\caption{The figure shows the behavior of $\rho$ with respect to $r$ for different values of $\beta$. We consider $\alpha=1,\, r_0=1$.}
\label{fig7}
\end{figure}

\begin{figure}[H]
\centering
\includegraphics[width=14.5cm,height=5cm]{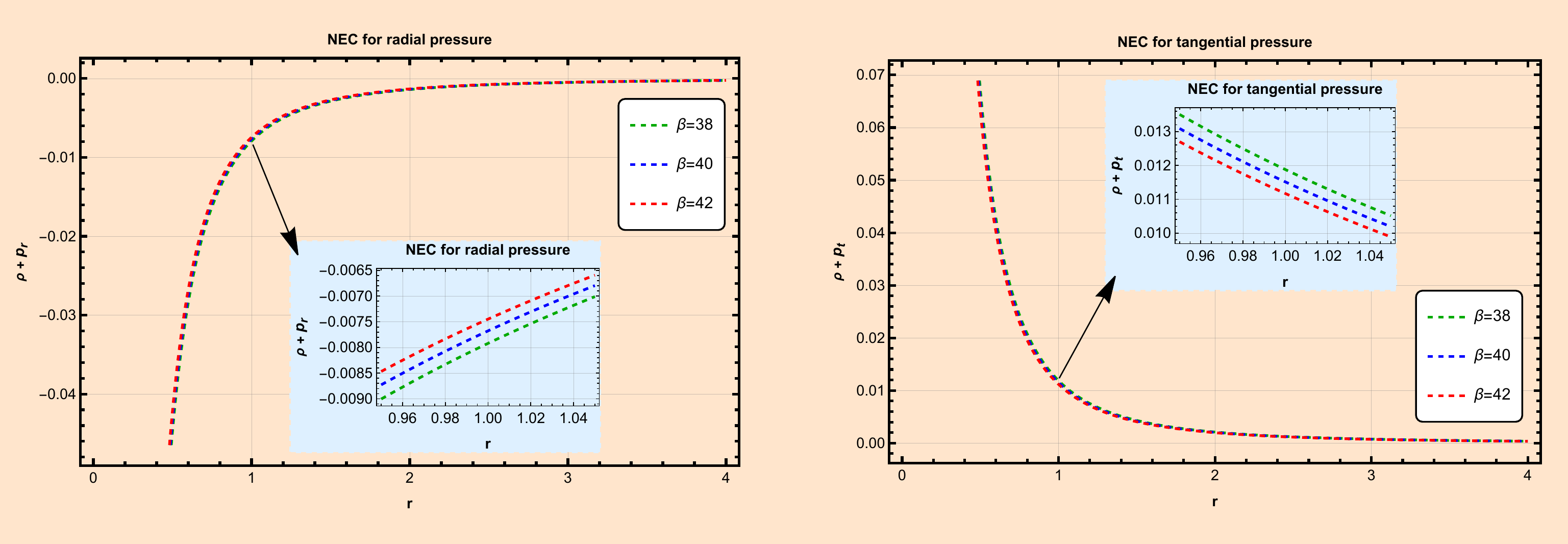}
\caption{Profile shows the behavior of NEC with respect to $r$ for different values of $\beta$. We consider $\alpha=1,\,r_0=1$.}
\label{fig8}
\end{figure}

\begin{figure}[H]
\centering
\includegraphics[width=14.5cm,height=5cm]{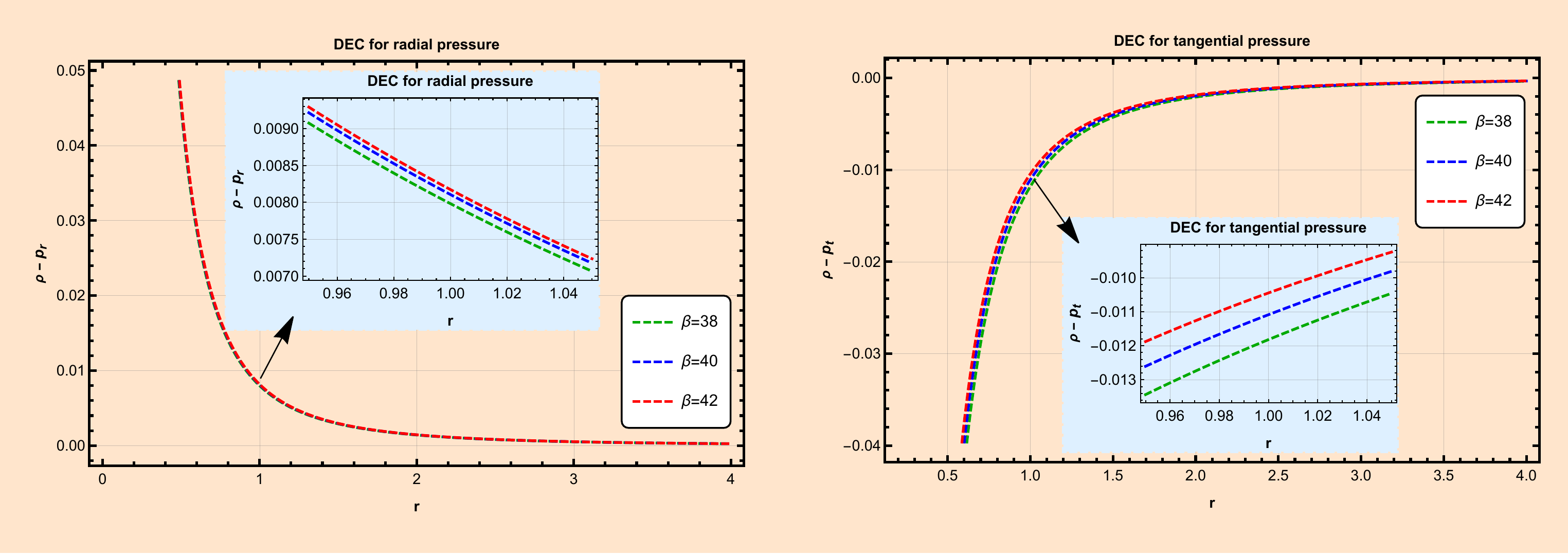}
\caption{Profile shows the behavior of DEC with respect to $r$ for different values of $\beta$. We consider $\alpha=1,\,r_0=1$.}
\label{fig9}
\end{figure}

\begin{figure}[H]
\centering
\includegraphics[width=6.5cm,height=4.7cm]{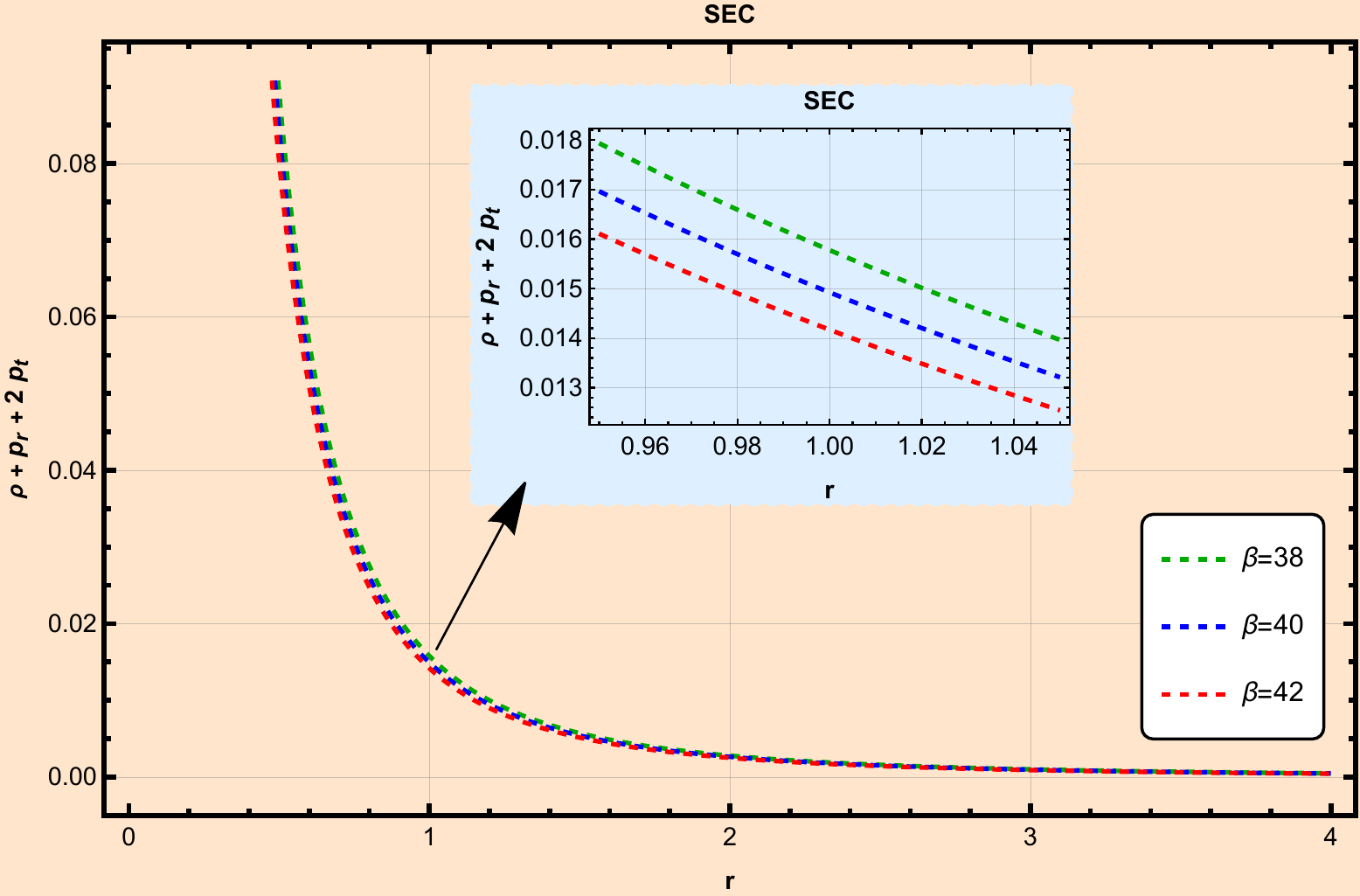}
\centering
\caption{Profile shows the behavior of SEC with respect to $r$ for different values of $\beta$. We consider $\alpha=1,\,r_0=1$.}
\label{fig10}
\end{figure}

Taking Eqs. \eqref{2c} and \eqref{35}, we derive the following set of energy conditions\\
$\bullet$ \textbf{NEC :} $\rho+p_r=-\frac{1}{2 (\beta +8 \pi ) r^2 \sqrt{\frac{r}{r_0}}}$ and $\rho+p_t=\frac{3}{4 (\beta +8 \pi ) r^2 \sqrt{\frac{r}{r_0}}}$,\\
$\bullet$ \textbf{DEC :} $\rho-p_r=\frac{36 \pi -5 \beta }{6 (4 \pi -\beta ) (\beta +8 \pi ) r^2 \sqrt{\frac{r}{r_0}}}$ and $\rho-p_t=\frac{5 \beta +12 \pi }{12 (4 \pi -\beta ) (\beta +8 \pi ) r^2 \sqrt{\frac{r}{r_0}}}$,\\
$\bullet$ \textbf{SEC :} $\rho+p_r+2p_t=-\frac{2 \beta }{3 (4 \pi -\beta ) (\beta +8 \pi ) r^2 \sqrt{\frac{r}{r_0}}}$\,,\\
which are described in details in Figs. \ref{fig7} - \ref{fig10}. As in our first example, we observe that the energy density is always positive, which means that WEC is satisfied. Again we also observe that NEC is violated by radial pressure and satisfied by the tangential pressure, unveiling the presence of strange matter in the wormholes throats. Moreover, DEC is satisfied by the radial pressure and violated by the tangential pressure for $r<1$. Furthermore, SEC is satisfied, so WEC and SEC allow the existence of stable traversable wormholes with strange matter on their throats. We also highlight that these constraints over the energy conditions are observed for $\tilde{\rho}$, $\tilde{p}_r$, and $\tilde{p}_t$ too.  Moreover, imposing that the parameter $B$, and the energy density $\rho$ are positive defined, besides the required energy conditions, we find that $\beta>12\pi$ for shape functions \eqref{33} and \eqref{35(a)}, if we consider $\alpha=1,\, \gamma=0.5,\, \text{and}\,\,r_0=1$.



\section{Equilibrium Conditions}
\label{sec5}
In this section, we consider the generalized Tolman-Oppenheimer-Volkov (TOV) equation \cite{Oppenheimer,Gorini,Kuhfittig} to find the stability of our wormholes in $f(Q,T)$ gravity. The generalized TOV equation can be written as
\begin{eqnarray}\label{38}
\frac{\varpi^{'}}{2}(\tilde{\rho}+\tilde{p_r})+\frac{d \tilde{p_r}}{dr}+\frac{2}{r}(\tilde{p_r}-\tilde{p_t})=0,
\end{eqnarray}
where $\varpi=2\phi(r)$.\\
Due to anisotropic matter distribution, the hydrostatic, gravitational, and anisotropic forces are defined as follows
\begin{eqnarray}\label{39}
F_h=-\frac{d \tilde{p_r}}{dr}, ~~~~~F_g=-\frac{\varpi^{'}}{2}(\tilde{\rho}+\tilde{p_r}), ~~~~F_a=\frac{2}{r}(\tilde{p_t}-\tilde{p_r}).
\end{eqnarray}
The stability of the wormhole solutions, imposes the constraint $F_h+F_g+F_a=0$. Since in this study, we have assumed the redshift function $\phi(r)=constant$, the gravitational contribution $F_g$ is null, so the stability constraint is rewritten as
\begin{equation}
\label{40}
F_h+F_a=0.
\end{equation}
Then, by taking Eqs. \eqref{25}, \eqref{24a14}, \eqref{24a15}, \eqref{2c}, \eqref{33}, and \eqref{39}, we get the following equations for the hydrostatic and anisotropic forces 
\begin{equation}\label{41}
F_h=\frac{r_0 (4 r_0 \gamma -3 (\gamma +1) r)}{8 \pi  r^5}\,,
\end{equation}
\begin{equation}\label{42}
F_a=\frac{r_0 (3 (\gamma +1) r-4 r_0 \gamma )}{8 \pi  r^5}\,.
\end{equation}
Here we considered the shape function-1. Moreover, by combining the shape function-2, together with Eqs. \eqref{25}, \eqref{24a14}, \eqref{24a15}, \eqref{2c}, \eqref{35(a)}, and \eqref{39}, we yield to
\begin{equation}\label{43}
F_h=-\frac{5}{16 \pi  r^3 \sqrt{\frac{r}{r_0}}}\,,
\end{equation}
\begin{equation}\label{44}
F_a=\frac{5}{16 \pi  r^3 \sqrt{\frac{r}{r_0}}}\,,
\end{equation}
as the equations for the hydrostatic and anisotropic forces, respectively. We can verify that the  hydrostatic and the anisotropic forces are independent of $\omega$ and $\beta$ for both shape functions. 
\begin{figure}[H]
\centering
\includegraphics[width=16cm,height=5cm]{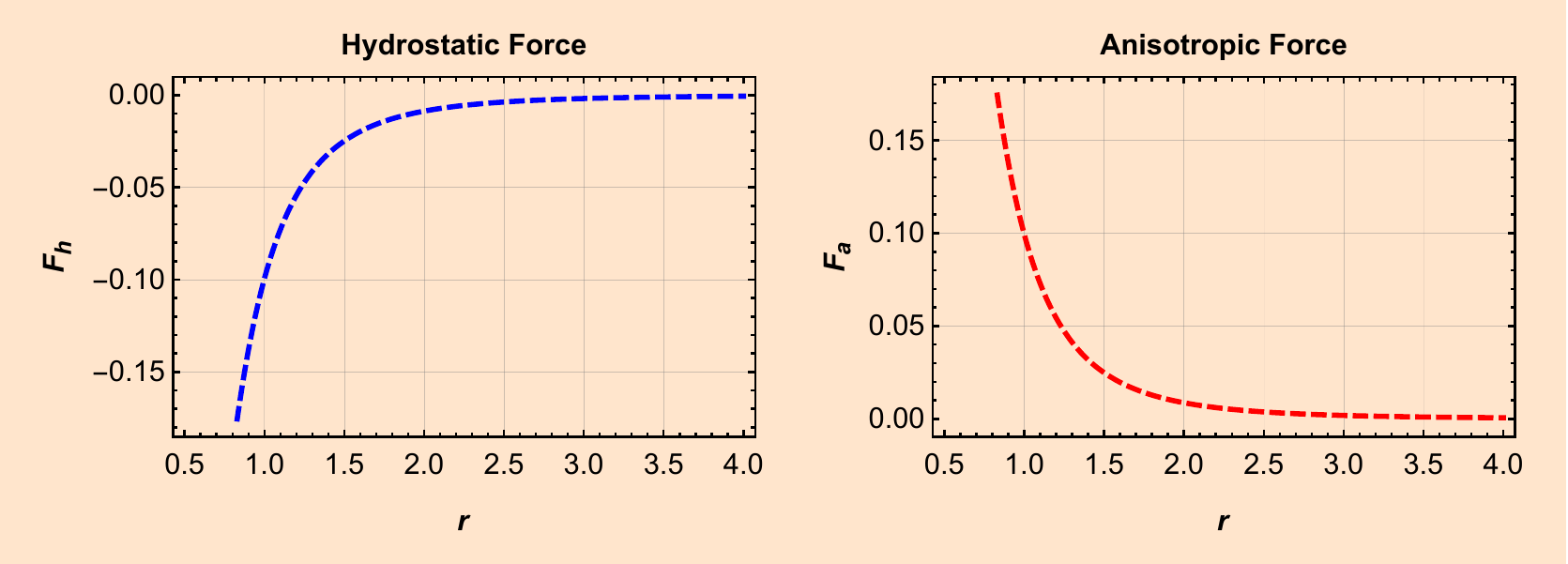}
\includegraphics[width=7.3cm,height=4.7cm]{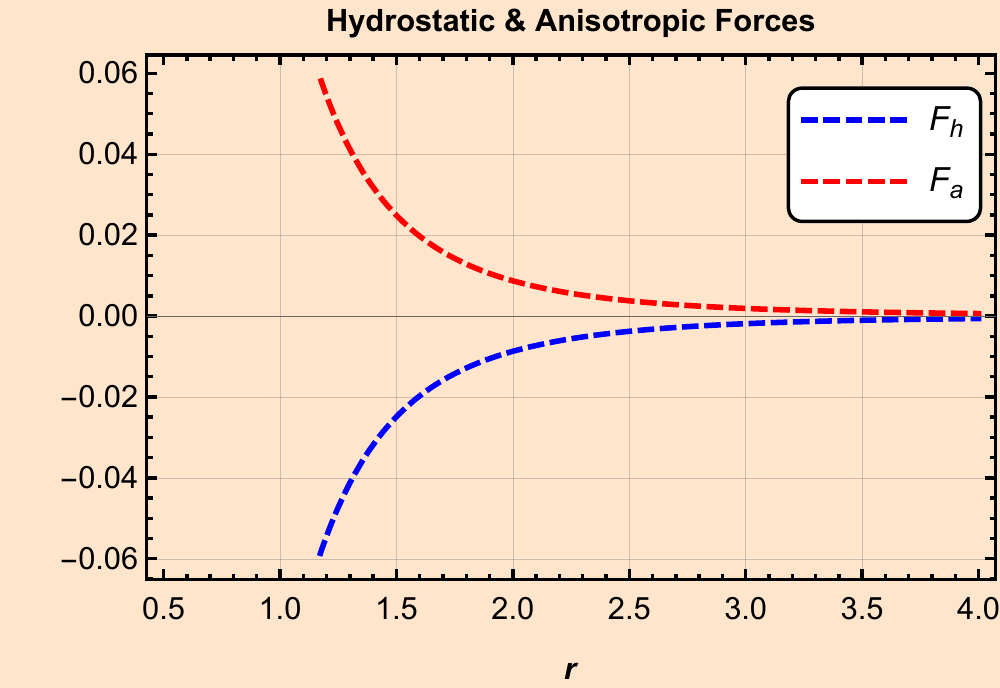}
\caption{This profile shows the behavior of hydrostatic forces and anisotropic forces for shape function-1. We consider $\gamma=0.5,\,r_0=1$.}
\label{fig11}
\end{figure}
\begin{figure}[H]
\centering
\includegraphics[width=14.5cm,height=5cm]{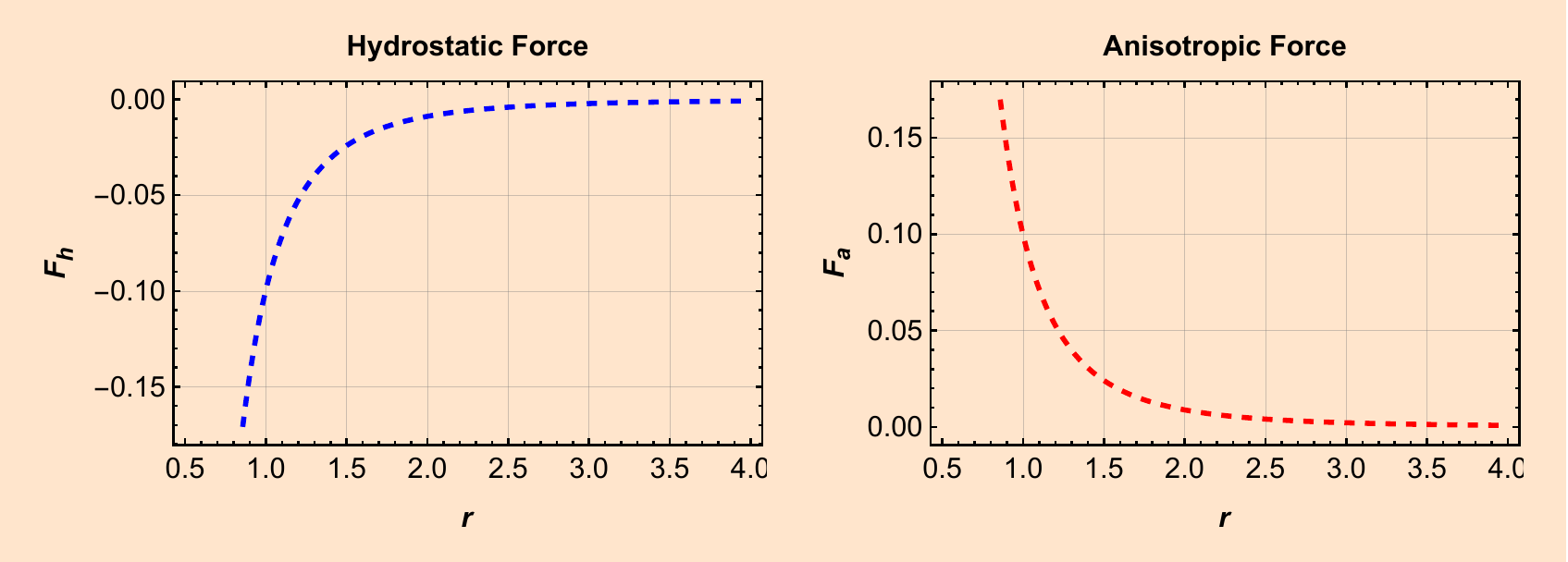}
\includegraphics[width=7.2cm,height=4.6cm]{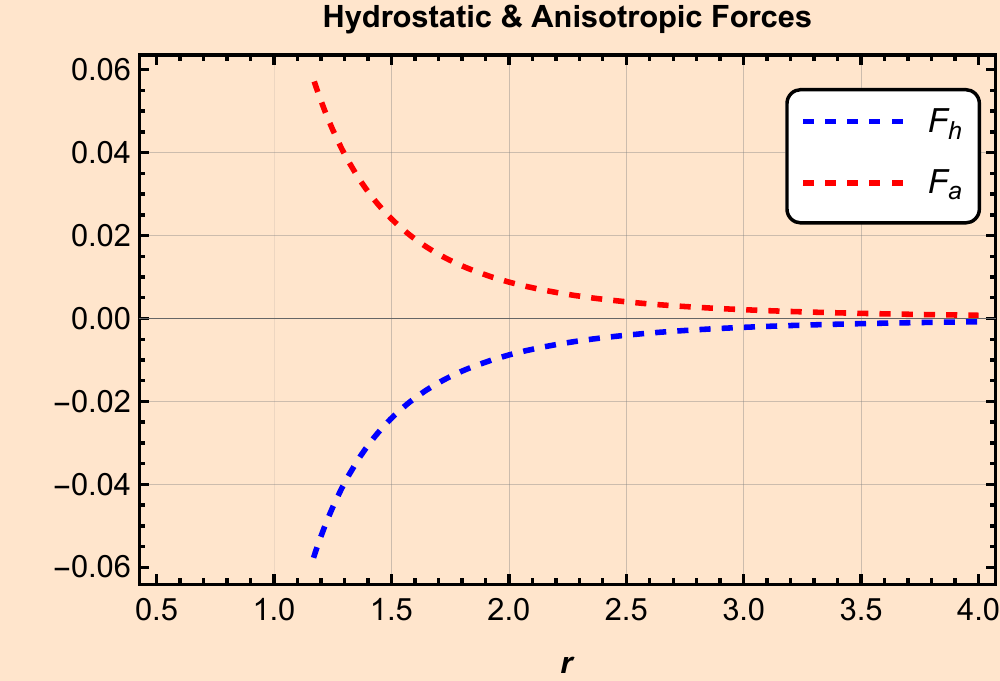}
\caption{This profile shows the behavior of hydrostatic forces and anisotropic forces for shape function-2. We consider $r_0=1$.}
\label{fig12}
\end{figure}
The graphics for the hydrostatic and the anisotropic forces derived from shape functions 1 and 2 are depicted in Figs. \ref{fig11} and \ref{fig12}. It can be observed that these forces show the same intensity but are opposite to each other. These balancing between the forces unveil that our wormholes are stable. 

\section{Final Remarks}
\label{sec6}

This investigation presents a new route to find the viability of wormholes solutions in $f(Q,T)$ gravity for a radial dependent bag parameter. In our discussions we derived different families of wormholes solutions using two different shape functions. Both shape functions are largely used in the recent literature on wormhole solutions for alternative gravity theories.

The energy conditions were derived through constraints imposed by the Raychaudhuri using the embedding procedure introduced by Mandal et al. \cite{Mandal/2020}. We used the energy conditions together with the equation of state for the MIT Bag Model, and with bounds over $\omega$ to derive interesting wormholes families for different values of our free parameter $\beta$. Such a procedure enable us to find wormholes solutions that satisfy at least two energy conditions for positive energy densities. 

Both shape functions unveiled wormhole solutions with positive energy densities satisfying WEC and SEC. The NEC and DEC energy conditions were partially violated for $r<1$, corroborating with the presence of strange matter at the wormholes throats.  The relation between the strange matter and the partial violation of NEC can be observed by analyzing Eq. \eqref{32}. There, by constraining that $\rho$, $\omega$, and $B$ are positive defined and that $B>\rho$ inside of the wormhole throat, we verify that NEC is violated. Therefore, our approach presents an interesting scenario with the possibility of stable traversable wormhole solutions formed by strange matter which satisfy SEC and WEC. We also performed a deep stability analysis of our solutions using the Tolman-Oppenheimer-Volkov (TOV) equation.

The methodology here presented can be applied to several other theories of gravity, opening a new path to obtain wormhole solutions with strange matter content, and constraining them through energy conditions. It would be interesting to derive wormhole solutions that obey the equation of state of the MIT Bag Model in noncommutative geometry, following the approach presented by Hassan et al. in \cite{Hassan/2021}. We hope to report on some of these ideas in near future. 

\section*{Data Availability Statement}

There are no new data associated with this article.

\acknowledgments 
M.T. acknowledges University Grants Commission (UGC), New Delhi, India, for awarding National Fellowship for Scheduled Caste Students (UGC-Ref. No.: 201610123801). JRLS would like to thank CNPq (Grant nos. 420479/2018-0, and 309494/2021-4), and PRONEX/CNPq/FAPESQ-PB (Grant nos. 165/2018, and 0015/2019) for financial support. JNA would like to thank CNPq PIBIC for financial support. P.K.S. acknowledges National Board for Higher Mathematics (NBHM) under Department of Atomic Energy (DAE), Govt. of India for financial support to carry out the Research project No.: 02011/3/2022 NBHM(R.P.)/R\&D II/2152 Dt.14.02.2022. We are very much grateful to the honorable referees and to the editor for the illuminating suggestions that have significantly improved our work in terms of research quality, and presentation.


\end{document}